\documentstyle[12pt,titlepage]{article}
\newlength{\extraspace}
\newlength{\extraspaces}
\catcode`\@=11

\def\numberbysection{\@addtoreset{equation}{section}
\def\theequation{\arabic{section}.\arabic{equation}}}

\newcommand{\be}{\begin{equation}
\addtolength{\abovedisplayskip}{\extraspaces}
\addtolength{\belowdisplayskip}{\extraspaces}
\addtolength{\abovedisplayshortskip}{\extraspace}
\addtolength{\belowdisplayshortskip}{\extraspace}}
\newcommand{\ee}{\end{equation}}
\newcommand{\ba}{\begin{eqnarray}
\addtolength{\abovedisplayskip}{\extraspaces}
\addtolength{\belowdisplayskip}{\extraspaces}
\addtolength{\abovedisplayshortskip}{\extraspace}
\addtolength{\belowdisplayshortskip}{\extraspace}}
\newcommand{\ea}{\end{eqnarray}}
\newcommand{\nonu}{\nonumber \\[.5mm]}
\newcommand{\tr}{\, {\rm tr} \,}

\newcommand{\VEV}[1]{\left\langle {#1} \right\rangle}
\newcommand{\Z}{\bf Z}
\newcommand{\R}{\bf R}

\newcommand{\tg}{\tilde{g}}

\newcommand{\tbeta}{\tilde{\beta}}
\newcommand{\betaeff}{\beta^{\mbox{\scriptsize eff}}}

\begin{document}
\begin{titlepage}
\addtolength{\baselineskip}{.5mm}
\thispagestyle{empty}
\begin{flushright}
TIT/HEP--219 \\
NUP-A-93-8 \\
hep-th/9307154\\
April, 1993
\end{flushright}
\begin{center}
{\large{\bf
Renormalization Group Approach \linebreak
to Matrix Models and Vector Models
\footnote{An expanded and updated version of the talk presented
by S.~Higuchi
at Yukawa Institute for Theoretical Physics workshop ``Quantum Gravity''
(24--27, November 1992).}
}} \\[6mm]
%%%%%%%%%%%%%%%%%%%%%%%%%%%%%%%%%%%%%%%%%%%%%
%
{\sc Saburo Higuchi}
\footnote{{\tt e-mail: hig@phys.titech.ac.jp}, JSPS fellow}\\[2mm]
{\it Department of Physics, Tokyo Institute of Technology, \\
Oh-okayama, Meguro, Tokyo 152, Japan} \\[2mm]
{\sc Chigak Itoi}
\footnote{\tt e-mail: itoi@phys.titech.ac.jp} \\[2mm]
{\it Department of Physics and Atomic Energy Research Institute, \\
College of Science and Technology, Nihon University, \\
Kanda Surugadai, Chiyoda, Tokyo 101, Japan} \\[2mm]
and \\[2mm]
{\sc Norisuke Sakai}
\footnote{\tt e-mail: nsakai@phys.titech.ac.jp} \\[2mm]
{\it Department of Physics,
Tokyo Institute of Technology, \\
Oh-okayama, Meguro, Tokyo 152, Japan} \\[6mm]
{\bf Abstract}\\[3mm]
{\parbox{13cm}{\hspace{5mm}
The renormalization group approach is studied for large $N$ models.
The approach of Br\'ezin and Zinn-Justin is explained and examined
for matrix models.
The validity of the approach is
clarified by using the vector model as a similar and simpler example.
An exact difference equation is obtained which relates
free energies for neighboring values of $N$.
The reparametrization freedom in field space provides infinitely many
identities which
reduce the infinite dimensional coupling
constant space to
that of finite dimensions.
The effective beta functions give exact values for the fixed points
and the susceptibility exponents.
A detailed study of the effective renormalization group flow is
presented for cases with up to two coupling constants.
}}
\end{center}
\end{titlepage}
\setcounter{section}{0}
\setcounter{equation}{0}
%%%%%%%  Introduction  %%%%%%%%%%%%%%%%%%%%%%%%%%%%%%%%%%%%%%%%%%%
\section {Introduction}
There has been a remarkable progress recently in understanding
the two-dimensional quantum gravity.
There are two main motivations to study the two-dimensional quantum
gravity coupled to matter.
Firstly, it is precisely a string theory when the two-dimensional
space-time is regarded as the world sheet for the string.
Secondly, it provides a toy model for the quantum gravity in higher
dimensions such as four dimensions.
There are two approaches to study the two-dimensional quantum gravity.
First one is the matrix model which gives a discretized  version
of the two-dimensional quantum gravity \cite{BIPZ}, \cite{GRMI},
\cite{GMIL}.
The second one is the Liouville theory
as a continuum theory \cite{DDK}, \cite{BEKL}.

The matrix model is quite powerful in providing a nonperturbative
treatment.
Exact solutions of the matrix model \cite{BIPZ}--\cite{GMIL}
have been obtained for two-dimensional quantum gravity coupled to
minimal conformal matter with central charge $c \le 1$.
The result can also be understood by means of the continuum approach
using the conformal field theory
\cite{DDK}, \cite{BEKL}.
In both approaches, it has been very difficult to obtain results
 for two-dimensional quantum gravity coupled to
conformal matter with central charge $c \ge 1$.
In matrix model approach, it is easy to write down matrix
models for cases with $c \ge 1$ as well \cite{BRHI}.
However, these models are not solvable up to now \cite{DDSW}.
The numerical simulations suggest that it is not at all
obvious if a matrix model
candidate to describe a $ c>1 $ model has the
continuum description \cite{DFJ}.
Although several exact solutions of the matrix model have been
obtained, it is worth studying approximation schemes which
enable us to  understand  unsolved matrix models
 at least qualitatively, especially for the case of $c>1$.
In order to make use of such a scheme, however,
we first need to make sure
that the approximation method gives correct results for the
exactly solved cases.
\par
Recently Br\'ezin and Zinn-Justin have proposed a renormalization group
approach to the matrix model\cite{BRZJ}.
They drew an analogy between the matrix model and the critical
phenomena in statistical mechanics.
The matrix model possesses the double scaling limit \cite{GRMI}.
The limit specifies
how the coupling constant should approach to a
critical value as the
size of the matrix $N$ goes to infinity.
This limit can be regarded as a continuum limit where bare parameters
should  approach to a
critical value as the
cut-off of the theory goes to infinity.
The critical exponents determine how the coupling constant should be
tuned to reach the continuum limit.
Therefore the double scaling limit
of the matrix model may be regarded as coming from
the fixed point of the renormalization group flow.
They observed that a change  $N \rightarrow N+\delta N$ can
be compensated by a change of coupling
constants $g \rightarrow g+\delta g$  in order
to give the same continuum physics.
They needed to enlarge the coupling constant space
as in the Wilson's renormalization group approach \cite{WIKO}.
Consequences of their approach have been examined by several groups
\cite{ALDA}.
A similar approach has been advocated previously for
the $1/N$ expansion in a somewhat different context
such as the $1/N$ corrections or the string field theory \cite{CARL}.

In the case of the one-matrix model with $c\le 1$,
Br\'{e}zin and Zinn-Justin obtained reasonable results
 for the fixed point and the susceptibility exponents
in the first nontrivial approximation.
In order to demonstrate the validity of the renormalization
group approach, however, one should show that
the systematic improvement
of their approximate evaluation converges to the correct result.
To this end, it is important to study a model in which a
renormalization group equation can be derived exactly, even if
it is simpler than the matrix model.
The vector model has been proposed for a discretized
one-dimensional quantum gravity, in the same way as
the matrix model for a discretized
two-dimensional quantum gravity\cite{NIYO}--\cite{ZJ}.
\par

Recently we have analyzed the vector model by means of the
renormalization group approach and
have clarified its validity and meaning \cite{HIS}.
This paper is a more complete account of our analysis of the vector
model and some results on the matrix model.
For the vector model, we have obtained an exact difference equation
which relates the free energy $-\log Z_{N-2}(g)$
to the free energy $-\log Z_N(g - 2\delta g)$ with
slightly different values of coupling constants.
We have found that these coupling constant shifts $\delta g_k$ are
of order $1/N$ and occur in infinitely many coupling constants.
We also obtained infinitely many identities which
express the freedom to reparametrize the field space.
By using these identities, we can rewrite the flow
in the infinite dimensional coupling constant space
as an effective flow in the space of finite
number of coupling constants.
The resulting effective beta function determines the fixed
points and the susceptibility exponents.
The inhomogeneous
term in the effective renormalization group equation
serves to fix non-universal (analytic)
terms of the free energy.
To illustrate the procedure by an explicit example,
we analyze in detail the cases of one and two coupling constants which
give the first two
multicritical points $m=2$ and $3$.
We obtain the fixed points and the
susceptibility exponents which are in complete agreement with the exact
results.
\par
In Sect.~2, the renormalization group approach for large $N$ models
is described and our results on matrix models are given.
In Sect.~3, the exact difference equation is obtained for vector
models.
In Sect.~4, reparametrization identities are obtained and the effective
renormalization group equations are derived for the single coupling
constant case.
In Sect.~5, the case of the multi-coupling constant is worked out.
The two coupling constant case is studied in detail, and
the renormalization group flow is illustrated in the two-dimensional
space of coupling constants.
Sect.~6 is devoted to a discussion.

%%%%%%%%%%%%%%%%%%%%%%%%
\section {Renormalization Group Approach to Matrix Models}
We first recall the renormalization group approach for the
matrix model. The partition function $Z_N(g)$ of the matrix
model with a single coupling constant $g$ for the cubic
interaction
is defined by an
integral over an $N\times N$ hermitian matrix $\Phi$
\be
Z_N(g)= \int d^{N^2} \Phi \exp \left[-N\tr\left({1 \over 2}\Phi^2
+{g \over 3}\Phi^3 \right) \right].
\ee
%\footnote{
Here we consider the cubic interaction, while
    Br\'{e}zin and Zinn-Justin studied the case of the quartic
    one. It seems that the ${\Z}_2$-symmetry of the quartic
    potential is not particularly useful to perform the perturbation
    in higher orders.
    We take the cubic one because we need less calculation in
    that case.
%}
The matrix model gives the random triangulation
of two-dimensional surfaces \cite{BIPZ}.
The $1/N$ expansion of the free energy $F(N, g)$
\be
 F (N, g)= -{1 \over N^2}\log Z_N(g)
= \sum_{h=0}^{\infty} N^{-2h} f_h(g)
\ee
distinguishes the contributions $f_h$ from the surface with $h$
handles.
In the double scaling limit
\be
 N \rightarrow \infty,   \qquad g \rightarrow g_*,  \qquad
{\rm with}   \quad   N^{2 \over \gamma_1}(g - g_*)
\quad {\rm fixed,}
\label{eqn:doublescalinglimit}
\ee
the singular part of the free energy satisfies the scaling law
 \cite{DDK}
with the susceptibility exponent
$\gamma_0 + h \gamma_1$
linear in the number of handles
$h$
\ba
 f_h (g)_{sing} &\!\!\! = &\!\!\!
(g - g_*)^{2 - \gamma_0-\gamma_1 h} a_h + \cdots, \nonu
 F (N,g)_{sing} &\!\!\! = &\!\!\! (g - g_*)^{2 - \gamma_0}
f \bigl( N^{2 \over \gamma_1}(g - g_*) \bigr).
\ea
The functional form of $f$ can be determined by a nonlinear
differential equation (string equation) \cite{GRMI}.
The continuum limit is achieved at the critical point
$g \rightarrow g_*$ where the average number of triangles
diverges.
The double scaling limit suggests that we may draw an analogy
between $N^{2 \over \gamma_1}$ and the momentum cut-off $\Lambda^2$
\ba
 g - g_* \rightarrow 0 \quad \quad & \Leftrightarrow &
 \quad \quad n_{\rm triangles} \rightarrow \infty \nonu
N^{2 \over \gamma_1} \rightarrow \infty \quad \quad & \Leftrightarrow  &
 \quad \quad a \rightarrow 0
\ea
where the lattice spacing of the random surface is denoted as $a$
and $\Leftrightarrow$ indicates the conjugate relation.
There is also an analogy between the scaling variables to be fixed
in the double scaling limit and the renormalized variables in the
continuum limit
\be
 N^{2 \over \gamma_1} (g - g_*)  \quad \quad
   \Leftrightarrow   \quad \quad
 a \cdot n_{\rm triangles}.
\ee

It has been proposed that the free energy $F(N,g)$ of the matrix model
satisfies the following renormalization group equation
 \cite{BRZJ}
\be
\biggl[N{\partial \over \partial N}
-\beta (g) {\partial \over \partial g}
+\gamma(g)\biggr] F (N, g)=r(g),
\label{eqn:matrixrge}
\ee
where $\beta(g)$ is called the beta function.
The anomalous dimension and the inhomogeneous term are
denoted as $\gamma(g)$ and $r(g)$ respectively.
A fixed point $g_*$ is given by a zero of the beta function.
The exponents of the
double scaling limit can be given by the derivative of the
beta function at the fixed point
\begin{eqnarray}
\gamma_0 = 2- \frac{\gamma(g_*)}{\beta'(g_*)},\\
\gamma_1 = \frac{2}{\beta'(g_*)}.
  \label{eqn:matrixexponent}
\end{eqnarray}

To obtain
the beta function, we integrate over a part of degrees of freedom of the
$(N+1)\times (N+1)$ matrix. \cite{BRZJ}, \cite{CARL}.
We parametrize the $(N+1)\times(N+1)$ matrix $\Phi_{N+1}$ as
\begin{equation}
  \Phi_{N+1} = \left(
  \begin{array}{cc}
    \Phi_N    & v \\
    v^\ast    & \alpha\\
  \end{array}
  \right),
\end{equation}
where $v$ is a complex $N$-vector and $\alpha$ is a real
variable.  Then we integrate over $v$ in the first order of the
perturbative expansion to obtain
\begin{equation}
  Z_{N+1}(g) = ( \lambda_N(g) )^{N^2} Z_{N} ( g + \delta g),
\end{equation}
where
\begin{eqnarray}
  \delta g & = & \frac{1}{N} \frac{11 g^3 -g }{2} + O(N^{-2}),\\
  \lambda  & = & 1 + \frac{1}{N} \frac{3g^2 -1}{2}    + O(N^{-2}).
\end{eqnarray}
We can read off $\beta,\gamma$ and $r$ in (\ref{eqn:matrixrge}) as
\begin{eqnarray}
  \beta(g) & = & \frac{1}{2}(11 g^3 -g),\\
  \gamma(g) & = & 2,\\
  r(g) & = & \frac{1}{2}(3g^2 -1).
\end{eqnarray}
Here we have used the assumption that there is no contribution
to these functions from the integration over $\alpha$, which
needs to be reexamined for our cubic potentials.

We find a zero of the beta function at $g_\ast=-1/\sqrt{11}=-0.302\ldots$
and identify this with the critical coupling constant.
Using (\ref{eqn:matrixexponent}), the critical exponents are
given as
\begin{equation}
  \gamma_0 = 0, \;  \gamma_1 = 2.
\end{equation}
Compared to the exact result
\begin{equation}
  g_\ast = - (432)^{-1/4} = -0.219\ldots , \;
  \gamma_0 = -1/2, \;   \gamma_1 = 5/2,
\end{equation}
this approximation seems to be reasonable. This situation is similar
to the case of the quartic potential studied by Br\'{e}zin
and Zinn-Justin \cite{BRZJ}.
In fact, the value of the critical exponent at this first nontrivial
order is precisely identical to their result.

In the first order approximation above,
induced interactions $\tr\Phi^4$ and $\tr\Phi^5$ of order $g^4$ and $g^5$
has been ignored consistently.
To perform the next order calculation,
we need to enlarge the coupling constant space and start with the
partition function
\begin{equation}
Z_N(g)= \int d^{N^2} \Phi \exp \left[-N\tr\left(
\frac{1}{2}\Phi^2 + \frac{g}{3}\Phi^3  +\frac{h}{4} \Phi^4+
\frac{f}{5} \Phi^5
\right) \right].
\end{equation}
Actually it turns out that, if we assume that $f \sim g^4, h \sim g^5$,
it is sufficient to include these two coupling constants
for the consistent second order calculation.
We obtain the beta functions
\begin{eqnarray}
\beta_g(g,h,f) & = & -\frac{1}{2}(g - 11g^3 + 18 g h - 6 f),\\
\beta_h(g,h,f) & = & -h -g^4,\\
\beta_f(g,h,f) & = & -\frac{1}{2}(3f-2g^5).
\end{eqnarray}
We find a common zero at
\begin{eqnarray}
g_\ast & = & -\left(\frac{-11+\sqrt{209}}{44} \right)^{1/2}
                   =-0.280\ldots,  \\
h_\ast & = & - g_\ast^4  =-0.00617\ldots,  \\
f_\ast & = & - \frac{2}{3} g_\ast^5    =-0.00115\ldots.
\end{eqnarray}
The scaling matrix at the fixed point
\begin{equation}
\left.
\left(
  \begin{array}{ccc}
\partial\beta_g/\partial g & \partial \beta_g /\partial h
                                 & \partial \beta_g /\partial f \\
\partial\beta_h/\partial g & \partial \beta_h /\partial h
                                 & \partial \beta_h /\partial f \\
\partial\beta_f/\partial g & \partial \beta_f /\partial h
                                 & \partial \beta_f /\partial f \\
  \end{array}
\right)
\right|_{g=g_\ast,h=h_\ast,f=f_\ast}
\end{equation}
exhibits three eigenvalues $1,-1.10\ldots, -1.55\ldots$.
Substituting the positive one into (\ref{eqn:matrixexponent}),
we obtain the critical exponents
\begin{equation}
 \gamma_0 = 0, \;   \gamma_1 = 2.
\end{equation}
Though approximate value for the critical coupling $g_\ast$ improves
slightly, those for the critical exponents show no improvement.
Before performing more elaborate calculation for the matrix
model, we find it
more illuminating to study the case of the vector model where we
can clarify the situation more fully.

%%%%%%%%%%%%%%%%%%%%%%%%
\section{Renormalization Group Approach to Vector Models}
The partition function of the $O(N)$ symmetric vector model is given by
\begin{equation}
  Z_N(g) = \int d^N \phi \;
  \exp \left[ -N \sum_{k=1}^{\infty} \frac{g_k}{2k} (\phi^2)^k\right],
  \label{eqn:vector_action}
\end{equation}
where $\phi$ is an $N$ dimensional real vector
\cite{NIYO} -- \cite{ZJ}.
Here we introduce infinitely many coupling constants $g_k$,
since we need all possible induced interactions after a
renormalization group transformation even if we start with a few
coupling constants only.
The $1/N$ expansion of the logarithm of the partition function
gives contributions from $h$ loops as terms with $N^{1-h}$.
The vector model  has the double scaling limit
$N \rightarrow \infty $ with $N^{1/\gamma_1}(g-g_\ast)$ fixed,
where
the singular part of the free energy satisfies the scaling law
\cite{NIYO} -- \cite{ZJ}
\be
 -\log \left.\biggl[{Z_N(g) \over Z_N(g_1=1,g_k=0 \; (k\geq2))}
 \biggr]\right|_{sing}
= \sum_{h=0}^{\infty} N^{1-h}
(g - g_*)^{2 - \gamma_0-\gamma_1 h} a_h + \cdots.
\ee
One should note that the power of $1/N$ and the definition of the
susceptibility exponents are slightly different from the
matrix model.
Therefore the relation between these susceptibility exponents and the
derivative of the beta function becomes
\be
\gamma_1 = {1 \over \beta'(g_*)}, \qquad
\gamma_0 = 2- {\gamma(g_*) \over \beta'(g_*)}.
 \label{eqn:vectorgamma_and_beta}
\ee

In the spirit of the
approximation method of ref.\cite{BRZJ}, we can integrate over
the $(N+1)$-th component $\alpha$ of the  vector $\phi_{N+1}$
\ba
\phi_{N+1} &\!\!\! = &\!\!\! (\phi_N, \alpha), \nonu
  Z_{N+1}(g)
    &\!\!\! = &\!\!\! \int d^N \phi_N \; d\alpha \;
            \exp \left[ -(N+1) \sum_{k=1}^{\infty}
                 \frac{g_k}{2k} (\phi_N^2+\alpha^2)^k\right].
\ea
Neglecting higher order terms in $1/N$, we obtain
\be
\frac{Z_{N+1}(g) }{Z_{N+1}(g_1=1,g_k=0\;(k\geq2))}
= \int d^N \phi_N
  \exp \left[ -N \sum_{k=1}^{\infty} \frac{g_k+\delta g_k}{2k}
       (\phi_N^2)^k + O\left({1 \over N}\right) \right],
   \label{eqn:approxbeta}
\ee
where the shifts $\delta g_k$ of the coupling constants are found to be
\begin{equation}
  \sum_{k=1}^{\infty} \frac{g_k}{k} x^k
  + \log \left(\sum_{k=1}^{\infty} \frac{g_k}{g_1} x^{k-1}\right)
  =
  N \sum_{k=1}^{\infty} \frac{\delta g_k}{k}    x^k.
   \label{eqn:shift}
\end{equation}

Now we shall show that the shift
$\delta g_k$ can be evaluated exactly in the vector model
without using approximation methods such as described above.
To demonstrate that the result is exact, we start with the partition
function $Z_{N-2}(g)$ .
After integrating over angular coordinates in ${\R}^{N-2}$,
we perform a partial integration in the radial coordinate
$x=\phi^2$
\begin{eqnarray}
  Z_{N-2}(g)
    &\!\!\! = &\!\!\! \frac{\pi^{\frac{N}{2}-1}}{\Gamma(\frac{N}{2}-1)}
          \int_{0}^{\infty} dx\;
       x^{N/2-2} \exp\left[-(N-2)\sum_{k=1}^{\infty}
          \frac{g_k}{2k}x^k\right] \nonu
    &\!\!\! = &\!\!\! \frac{\pi^{\frac{N}{2}-1}}{\Gamma(\frac{N}{2}-1)}
          \int_{0}^{\infty} dx\;
       x^{N/2-1} \left(\sum_{k=1}^{\infty} g_k x^{k-1}\right)
         \exp \left[-(N-2)\sum_{k=1}^{\infty} \frac{g_k}{2k}x^k\right].
   \label{eqn:radial_int}
\end{eqnarray}
Identifying the  right hand side with
$((N-2) g_1/ 2\pi) Z_{N}(g - 2 \delta g)$,
we obtain an exact difference equation for the
logarithm of the partition function
\begin{eqnarray}
\lefteqn{[(-\log Z_{N}(g))-(-\log Z_{N-2}(g))]
- \log\frac{(N-2) g_1}{2\pi}}
 \nonumber \\
 &\!\!\! =&\!\!\! - [ ( -\log Z_{N} ( g - 2 \delta g))
 - ( -\log Z_{N}(g))].
  \label{eqn:difference_equation}
\end{eqnarray}
We find that the shifts $\delta g_k$ of the coupling constants are
exactly identical to the result (\ref{eqn:shift})
of the approximate evaluation.
We would like to stress
that no approximation is employed to obtain
eq. (\ref{eqn:difference_equation}).
Therefore we can infer that
the vector model offers an example to justify the approximate evaluation
method of ref.\cite{BRZJ}.
\par
In the $N \rightarrow \infty$ limit, we can obtain a
differential equation from the exact
difference equation (\ref{eqn:difference_equation})
\begin{equation}
  \frac{\partial}{\partial N}( -\log Z_{N} ( g ))
        - \frac{1}{2} \log \frac{Ng_1}{2\pi}
  =
  \sum_{k=1}^{\infty}  \delta g_k \frac{\partial}{\partial  g_k}
( -\log Z_{N} ( g )).
  \label{eqn:differential_equation}
\end{equation}
One can bring the quadratic
term in the potential to the standard form $\phi^2/2$
since $g_1$ can be absorbed
by a rescaling $g_1 \phi^2 \rightarrow \phi^2$. We have
\begin{equation}
 Z_N(g_1,g_2,g_3,\ldots) = g_1^{-N/2} Z_N(1, g_2/g_1^2,g_3/g_1^3,\ldots).
\label{rescalingid}
\end{equation}
Therefore it is convenient to use the
rescaled coupling constants $\tilde g_k$ together with $g_1$
as independent coupling constants
\begin{equation}
 \tg_k = g_k / g_1^k.
\end{equation}
We shall define the free energy for the vector model
\begin{equation}
F(N,\tg) = - \frac{1}{N} \log Z_N(g)
- \frac{1}{2} \log \frac{Ng_1}{2\pi}.
\end{equation}
If we use $g_1, \tg_2, \tg_3, \cdots$  as independent coupling
constants, we find from the rescaling identity (\ref{rescalingid})
that the partition function $Z_N$ depends on $g_1$ only through the
factor $g_1^{-N/2}$.
Therefore, the free energy $F(N, \tilde g)$ is independent of
$g_1$ and is a function of $\tg_k \; (k \geq 2)$ only.
\par
We denote the partial derivatives with respect to
$g_1, \tilde g_2, \tilde g_3, \cdots$ by $|_{\tilde g}$, and
those with $g_1, g_2, g_3, \cdots$
by $|_{g} $
\begin{equation}
 \left.\frac{\partial}{\partial g_1}\right|_g
=\left.\frac{\partial}{\partial g_1}\right|_{\tg}
   - \sum_{k=2}^{\infty} k \frac{\tg_k}{g_1}
               \left.   \frac{\partial}{\partial    \tg_k}\right|_{\tg},
\qquad
 \left.\frac{\partial}{\partial g_k}\right|_g
= \frac{1}{g_1^k} \left. \frac{\partial}{\partial \tg_k}\right|_{\tg} .
\end{equation}
Thus we obtain a renormalization group equation for the free
 energy $F$
\begin{equation}
\left[N \frac{\partial}{\partial N}
  -
  \sum_{k=2}^{\infty}
N \left( \frac{\delta g_k}{g_1^k} - \frac{\delta g_1}{g_1} k \tg_k\right)
\left.     \frac{\partial}{\partial \tg_k} \right|_{\tg}
 +1   \right]F(N,\tg)
= N \delta g_1 \frac{1}{2g_1}
      - \frac{1}{2}   .
 \label{eqn:renorm_all_coupling}
\end{equation}
Eq.(\ref{eqn:renorm_all_coupling}) shows that the anomalous
dimension is  given by
\be
  \gamma(\tg) =1,
\label{eqn:anomalousdim}
\ee
which implies a relation between two susceptibility exponents
\be
  \gamma_0 + \gamma_1 = 2.
\ee
We read off the beta functions in the rescaled coupling
constants as
\begin{equation}
\tbeta_k(\tg)
=N \left(
     \frac{\delta g_k}{g_1^k} - \frac{\delta g_1}{g_1} k
     \tilde{g_k}
     \right).
   \label{eqn:rescaled_beta}
\end{equation}
These beta functions can be evaluated
explicitly using eq.(\ref{eqn:shift}) as
\begin{eqnarray}
  \tbeta_2(\tg) &\!\!\! = &\!\!\! -\tg_2 - 3 \tg_2^2 + 2 \tg_3, \nonu
  \tbeta_3(\tg) &\!\!\! = &\!\!\! -2\tg_3
  + \tg_2^3 -6\tg_2 \tg_3 + 3 \tg_4, \nonu
  \tbeta_4(\tg) &\!\!\! = &\!\!\!-3\tg_4 - \tg_2^4 + 4\tg_2^2\tg_3 -
  2\tg_3^2 - 8 \tg_2 \tg_4 + 4 \tg_5,  \nonu
                &\!\!\! \vdots &\!\!\! .
  \label{eqn:explicit_beta}
\end{eqnarray}

Let us investigate the simultaneous zero of the beta functions
$\tilde \beta_k=0 \; (k\geq 2)$ in the spirit of ref.\cite{BRZJ}.
The first condition $\tilde \beta_2 = 0$ gives $\tilde g_3$ in terms of
$\tilde g_2$.
The second condition $\tilde \beta_3=0$ determines $\tilde g_4$
in terms of $\tilde g_2$ and $\tilde g_3$, and hence
in terms of $\tilde g_2$.
As can be seen in eq.(\ref{eqn:explicit_beta}),
$\tbeta_k(\tilde g)$ turns out to be a sum of $k \tilde g_{k+1}$
and a polynomial in $\tilde g_j, j\le k$.
Therefore we find that there always exists a solution of
$\tilde \beta_k=0 \; (k\geq 2)$
for each given value of the coupling constant $\tg_2$.
In the following, we shall show that this strange result
of the apparent existence of
the one-parameter family of fixed points is due to a misinterpretation
of the renormalization group flow.
\par

%%%%%%%%%%%%%%%%%%%%%%%%
\section{Reparametrization Identities}
The key observation is the ambiguity to identify the renormalization
group flow in the coupling constant space.
Though the above equation (\ref{eqn:renorm_all_coupling}) seems to
describe a renormalization group flow in the infinite dimensional
coupling constant space,
the direction of the flow is in fact ambiguous
because all the differential operators $(\partial/\partial\tilde{g_k})$
are not linearly independent as we will see shortly.
To see this, note that the partition function (\ref{eqn:vector_action})
is invariant under reparametrizations of the integration variable
$\phi$.
Since the model is $O(N)$ invariant, we can obtain new informations
only from reparametrizations of the radial coordinate $x=\phi^2$.
Since the radial coordinate should take values on the half real line,
we consider the most general reparametrization which keeps the
integration range $[0,\infty)$
\begin{equation}
 x = y \left( 1+ \sum_{j=0}^{\infty} \varepsilon_j y^j\right),
 \label{eqn:reparametrization}
\end{equation}
where $\varepsilon_j$'s are infinitesimal parameters.
Substituting (\ref{eqn:reparametrization})
in (\ref{eqn:radial_int}) and differentiating with respect to
$\varepsilon_j$'s,
we obtain a family of identities
\begin{eqnarray}
 &\!\!\! &\!\!\! L_j Z_N(g) = 0  \mbox{\ \ \ } \qquad ( j \geq 0), \\
 &\!\!\! &\!\!\! L_j = \sum_{\ell=j+1}^{\infty}g_{\ell-j}\ell
           \left.\frac{\partial}{\partial g_\ell}\right|_g
        - \left( 1+ \frac{2j}{N}\right)
          \left. j \frac{\partial}{\partial g_j}\right|_g
           + \frac{N}{2} \delta_{j,0}.
\end{eqnarray}
The differential operators $L_j$ constitute  half of the
Virasoro algebra.
We can show that this algebra is identical with the one found in
\cite{NIYO} and \cite{DVKO}.

It is more useful to use the rescaled coupling constants $\tilde g_k$.
The reparametrization identity corresponding to $\varepsilon_0$ is
nothing but the infinitesimal form of the rescaling identity
(\ref{rescalingid}) and reads
\be
\left.\frac{\partial}{\partial g_1}\right|_{\tg} F(N,\tg)  =  0.
\ee
In terms of the rescaled coupling constants,
the reparametrization identity corresponding to $\varepsilon_1$ is
given as
\be
-\frac{N+2}{2N}+
 \sum_{l=2}^{\infty}
\left\{\left(1+\frac{2}{N}\right)\tg_l+\tg_{\ell-1}\right\}
\left. \ell \frac{\partial}{\partial \tg_\ell}\right|_{\tg} F(N,\tg) = 0.
\label{eqn:firstreparametid}
\ee
The reparametrization identities corresponding to $\varepsilon_j$ is
given by
\be
\left.\left\{-\left(1+\frac{2j}{N}\right)
j \frac{\partial}{\partial \tg_j}
 + \sum_{l=j+1}^{\infty} \tg_{\ell-j}
  \ell \frac{\partial}{\partial \tg_\ell}
\right\}\right|_{\tg} F(N,\tg) = 0.
\label{eqn:jreparametid}
\ee
We see that derivatives of the free energy in terms of
infinitely many coupling constants $\tilde g_k$
are related by
infinitely many reparametrization identities.
Thus one can expect that only a finite number of derivatives are
linearly independent.
The exact difference equation (\ref{eqn:difference_equation})
combined with the reparametrization identities
(\ref{eqn:firstreparametid}) and (\ref{eqn:jreparametid})
constitute the complete set
of equations to characterize the renormalization group flow in our
approach.

To illustrate the use of the reparametrization identities,
we shall first take the case of a single coupling constant.
Let us consider a point in the coupling constant space
\be
(g_1, \tg_2, \tg_3, \tg_4, \ldots) = (g_1,\tg_2,0,0, \ldots).
\label{eqn:m2subspace}
\ee
At this point, the $j$-th identity relates
$\partial F / \partial \tg_{j+2}$ to
$\partial F / \partial \tg_{j+1}$ except for the case of $j=1$
where an extra constant term is present.
Therefore we can
express $\partial F / \partial \tg_k \; \; (k \geq 3)$ in terms
of $\partial F / \partial \tg_2 $ by
solving these reparametrization identities recursively
in the one-dimensional subspace (\ref{eqn:m2subspace}) of coupling
constants.
It is most convenient to organize the solution in powers of $1/N$.
We find explicitly at leading order
\begin{equation}
\frac{\partial F}{\partial \tg_k}
= B_k \frac{\partial F}{\partial \tg_2}  + R_k + O(N^{-1}),
\end{equation}
where
\begin{eqnarray}
B_k &\!\!\! = &\!\!\! \frac{(1-\Delta)^2}{16\sqrt{\Delta}}
         \left\{
       \left(\frac{2}{1-\sqrt{\Delta}}\right)^k
     - \left(\frac{2}{1+\sqrt{\Delta}}\right)^k
        \right\}           \frac{2}{k},\\
R_k &\!\!\! = &\!\!\! - \frac{1}{\sqrt{\Delta}}
         \left\{
        \left(\frac{2}{1-\sqrt{\Delta}}\right)^{k-2}
     - \left(\frac{2}{1+\sqrt{\Delta}}\right)^{k-2}
         \right\}           \frac{1}{2k},\\
\Delta &\!\!\! = &\!\!\! 4 \tg_2 + 1.
\end{eqnarray}
These solutions at leading order are sufficient to
eliminate $\partial F/\partial \tg_k  \; \; ( k \geq 3) $ in favor of
$\partial F/\partial \tg_2$ in (\ref{eqn:renorm_all_coupling}).
Thus we obtain a renormalization group equation with
the effective beta function $\betaeff(\tg_2)$
and the inhomogeneous term $r(\tilde g_2)$
\begin{equation}
\left[N\frac{\partial}{\partial N}
- \betaeff(\tg_2) \frac{\partial}{\partial \tg_2} + 1 \right]
 F(N,\tg_2) = r( \tg_2),
  \label{eqn:effective_rge}
\end{equation}
where
\begin{eqnarray}
\betaeff(\tg_2)
&\!\!\! = &\!\!\! \sum_{k=2}^{\infty} B_k
\tbeta_k(\tilde g_2, \tilde g_k=0,k\geq 3)
\nonu
&\!\!\! = &\!\!\! {1 \over 4}
\left[1-\Delta - {(1-\Delta)^2 \over 2\sqrt{\Delta}}
\log \left({1+\sqrt{\Delta} \over 1-\sqrt{\Delta}}\right)\right]
\nonu
&\!\!\! = &\!\!\! \frac{2}{3} \left( \tg_2 + \frac{1}{4}\right)
     - \frac{32}{15} \left(\tg_2 + \frac{1}{4}\right)^2
       + \cdots,
\label{eqn:effectivebeta}
\\
r(\tg_2)
&\!\!\! = &\!\!\!
\frac{N}{2} \delta g_1 - \frac{1}{2}
+ \sum_{k=2}^{\infty} R_k
\tbeta_k(\tilde g_2, \tilde g_k=0,k\geq 3)
\nonu
&\!\!\! = &\!\!\! {1 \over 2\sqrt{\Delta}}
\left[\left({1+\sqrt{\Delta} \over 2}\right)^2
\log \left({1+\sqrt{\Delta} \over 2}\right)
-\left({1-\sqrt{\Delta} \over 2}\right)^2
\log \left({1-\sqrt{\Delta} \over 2}\right)\right]
\nonu
&\!\!\! = &\!\!\! \frac{1}{4} + \frac{1}{2} \log \frac{1}{2}
       + \frac{1}{3}\left( \tg_2 + \frac{1}{4}\right)
       + \frac{2}{15}\left(\tg_2 + \frac{1}{4}\right)^2 + \cdots.
\label{eqn:inhomogeneousterm}
\end{eqnarray}

The graphs of the effective beta function $\betaeff(\tg_2)$ and
the inhomogeneous term $r(\tg_2)$ are shown in figure 1.%~\ref{fig:1beta}
% figure 1
We find the beta function has two zeros $\tg_2 = -1/4$
and $\tg_2=0$.

The fixed point $\tg_2 = -1/4$ is an infrared-unstable fixed
point. By choosing the bare coupling constant around this point,
we obtain a non-trivial theory with the renormalized coupling
constant $-\infty < \tg_2 < 0$. Here we use the term
``infrared'' for the limit $N \rightarrow 0$ in analogy with the
ordinary field theory.

Furthermore, we can calculate the susceptibility exponents from
the derivative of the
effective beta function
by using (\ref{eqn:vectorgamma_and_beta}) and (\ref{eqn:anomalousdim})
\be
\gamma_1 = {1 \over \beta'(\tg_{2*})} = \frac{3}{2}, \qquad
\gamma_0 = 2- {\gamma(\tg_{2*}) \over \beta'(\tg_{2*})}
= \frac{1}{2}.
\ee
The fixed point and the susceptibility exponent are in complete
agreement with the exact results for the $m=2$ critical point of
the vector model corresponding to pure gravity \cite{NIYO} -- \cite{AMP}.

Another fixed point $\tg_2=0$ is an infrared-stable fixed point
or the ``gaussian'' fixed point, since
$\partial\betaeff/ \partial \tilde g_2=-1<0$.
Without particular tuning of
the coupling constant we obtain a free theory corresponding to
this fixed point.

We observe that both the effective beta function
$\betaeff(\tg_2) $
and the inhomogeneous term $r(\tilde g_2)$ are analytic in $\tg_2$
around the fixed point $\tg_2=-1/4$. However, they become complex for
$\tilde g_2 > 0$, as shown in fig.1.

We can extract the complete information from the renormalization
group flow in our approach, namely the exact difference equation
and the reparametrization identities
by a systematic expansion of the free energy in powers of $1/N$
\begin{equation}
F(N,\tg_2) = \sum_{h=0}^{\infty} N^{-h} f_h(\tg_2).
\end{equation}
The free energy at leading order
$f_0$  satisfies an ordinary differential equation exactly,
\begin{equation}
  f_0(\tg_2) -
\betaeff(\tg_2)\frac{\partial f_0}{\partial \tg_2}(\tg_2)
 = r(\tg_2),
 \label{eqn:diff_leadingN}
\end{equation}
where the effective beta function and the inhomogeneous term are
those given in eqs.(\ref{eqn:effectivebeta}) and
(\ref{eqn:inhomogeneousterm}).
We see immediately that the general solution is given by a sum
of an arbitrary multiple of the solution of the homogeneous equation and
a particular solution of the inhomogeneous equation.
Since both the effective beta function $\betaeff(\tg_2) $
and the inhomogeneous term $r(\tilde g_2)$ are analytic in $\tg_2$
around the fixed point $\tg_2=-1/4$,
the singular behaviour of $f_0$
comes from the solution of the homogeneous equation.
More explicitly, we can write $f_0$ as the sum of two terms
$f_0(\tg_2)_{sing}$ and $f_0(\tg_2)_{analytic}$.
The former is singular around $\tg_2=-1/4$ and satisfies the
homogeneous equation, while
the latter is analytic around $\tg_2=-1/4$ and is a particular solution
of the inhomogeneous equation (See figure 2 ).%\ref{fig:leadingfreeenergy})
\begin{eqnarray}
f_0(\tg_2)_{sing} & = &
      \frac{1}{4}\log\frac{1+\sqrt{\Delta}}{1-\sqrt{\Delta}}
        - \frac{\sqrt{\Delta}}{2(1-\Delta)},\\
f_0(\tg_2)_{analytic} & = &
      \frac{1}{4}\log\frac{1-\Delta}{4}
        + \frac{1}{2(1-\Delta)}-\frac{1}{4}.
  \label{eqn:leadingfreeenergy}
\end{eqnarray}

It is important to notice that the singular term
corresponding to the continuum physics (the so-called universal
term) is specified by the beta function alone. Its
normalization, however, cannot be obtained from the
renormalization group equation. The analytic contributions are
determined by the inhomogeneous term and the effective beta
function.

The exact solution for $h=0$ is obtained by choosing the
normalization constant for the singular term $f_0(\tg_2)_{sing}$
to be unity.
\begin{eqnarray}
f_0(\tg_2) & = & f_0(\tg_2)_{sing} + f_0(\tg_2)_{analytic}, \nonumber \\
           & = & \frac{1}{2} \log \frac{1+\sqrt{\Delta}}{2}
                 +\frac{1}{2(1+\sqrt{\Delta})}-\frac{1}{4}.
\end{eqnarray}
We find that the singular part of the free energy acquires an imaginary
part for $\tilde g_2 >0$. this is due to the fact that the separation
of the singular and analytic parts is defined by the behavior
at $\tilde g_2=-1/4$. Both
$ f_0(\tg_2)_{sing}$ and $ f_0(\tg_2)_{analytic} $ become complex
after extrapolation beyond $\tilde g_2>0$.
However, the full free energy $f_0(\tg_2)$ is of course real for
$\tilde g_2>0 $, in accordance with the stability of the potential in
that parameter region.
On the contrary, the full free energy $f_0(\tg_2)$ acquires an imaginary
part in $\tilde g_2< -1/4$, which comes entirely from the singular part
$ f_0(\tg_2)_{sing}$.
This phenomenon corresponds to the instability of the potential in
this parameter region.
% figure 2

To obtain the full information for the free energy up to the order
$N^{-h}$, we should
write down the solution of the recursive reparametrization
identities (\ref{eqn:firstreparametid}) and (\ref{eqn:jreparametid})
up to the order $N^{-h}$ explicitly.
By inserting the solution into
the exact difference equation (\ref{eqn:difference_equation})
and by expanding it up to the power $N^{-h}$, we
obtain an ordinary differential equation for $f_h$
\be
  (1-h)f_h(\tg_2) -
\betaeff(\tg_2)\frac{\partial f_0}{\partial \tg_2}(\tg_2)
= r_h(\tg_2).
\ee
We see that the effective beta function is common to all $h$,
whereas the inhomogeneous terms $r_h$ depend on $h$.
Therefore we find the singular part
\footnote{ We have included the possible logarithmic terms and
  integer power terms into the analytic contribution, since they
  are non-universal. The singular contributions are universal}
of the free energy
is determined by the beta function up to a
normalization $a_h$
\ba
f_h(\tg_2) &\!\!\! = &\!\!\!
f_h(\tg_2)_{sing}+f_h(\tg_2)_{analytic}, \nonu
f_h(\tg_2)_{sing} &\!\!\! = &\!\!\!
(\tg_2 - \tg_{2*})^{2 - \gamma_0-\gamma_1 h} a_h + \cdots.
\ea
If we sum over the contributions from various $h$, we find that
the renormalization group equation determines the combinations of
variables appropriate to define the double scaling limit.
On the other hand, the functional form of the scaled variable is
undetermined corresponding to the undetermined normalization factor
$a_h$ for the singular terms of each $h$
\ba
F(N, \tilde g) &\!\!\! = &\!\!\!
\sum_{h=0}^{\infty} N^{-h}f_h(\tg_2)_{sing}
=\sum_{h=0}^{\infty} N^{-h}
(\tg_2 - \tg_{2*})^{2 - \gamma_0-\gamma_1 h} a_h \nonu
&\!\!\! = &\!\!\! (\tg_2 - \tg_{2*})^{2- \gamma_0}
f\bigl(N^{1/\gamma_1}(\tg_2-\tg_{2*})\bigr)_{sing}.
\ea

It has been known in the exact solution of the vector model
that a nonlinear differential equation
determines the functional form
$f\left(N^{1/\gamma_1}(\tilde g_2-\tg_{2*})\right)_{sing} $
of the scaled variables.
The equation is called the string equation,
or the $L_{-1}$ Virasoro constraint \cite{GRMI}, \cite{FKN}.
The above consideration means that the complete set of
our renormalization group flow equations, namely the exact difference
equation and the reparametrization identities, does not give
the $L_{-1}$ constraint for the universal singular terms.
This is in accord with our objective of the renormalization
group approach: to obtain at least fixed points and critical
exponents even if the exact solution is not available.
On the other hand, our equations give informations
on the non-universal analytic terms.
We would like to point out, however, that
we can determine the functional form of
$f\left(N^{1/\gamma_1}(\tilde g_2-\tg_{2*})\right)_{sing} $
if we know the boundary condition of the renormalization group equation.
Namely, we have only to know the functional form of
$F(N=N_0,\tg_2)$, the free energy with a fixed $N$, or
$F(N,\tg_2=\tg_2^0)$, the free energy with a fixed coupling constant.

%%%%%%%%%%%%%%%%%%%
\section{Renormalization Group Flow in the \sloppy Multi-Coupling
  Constants Space}
It is known that there is a series of critical points labeled by
an integer $m=2,3,4,\ldots$ in vector models\cite{NIYO}--\cite{AMP}.
The critical exponents of the $m$-th critical point are given by
\begin{eqnarray}
  \gamma_0 = 1 - 1/m, \ \
  \gamma_1 = 1 + 1/m.
\end{eqnarray}
It is known that the $m$-th multicritical points constitute a
submanifold with codimension $m-1$ in the
infinite-dimensional coupling constants space.
In other words, we have to tune $m-1$ parameters to achieve
the $m$-th multicritical point.

In the previous analysis, we have found only the fixed point
corresponding to the $m=2$ critical point.
It is natural that we have not found higher multicritical points,
since we have limited our consideration in the one-dimensional coupling
constant space.
We can extend the analysis to the more general situation of finitely
many coupling constants.
Let us take
\be
g_1 = 1, \qquad
\tilde g_k \not= 0 \quad ( 2 \le k \le m), \qquad
\tilde g_k = 0 \quad ( k \ge m+1).
\ee
In this subspace, each reparametrization identity involves
only finite number of derivatives.
Moreover, the $j$-th identity relates derivatives in terms of
coupling constants $\tilde g_{j+1}, \tilde g_{j+2}, \cdots,
\tilde g_{j+m}$.
Therefore we can solve this identity to rewrite
$\partial F / \partial \tg_{j+m}$ in terms of
$\partial F / \partial \tg_k \; \; (j+1 \leq k \leq j+m-1)$.
By successively using these identities,
we find that there are precisely the necessary
number of recursion relations to express
$\partial F / \partial \tg_k \; \; (k \geq m+1)$
in terms of $\partial F / \partial \tg_k \; \; (2 \leq k \leq m)$.
Consequently, we can reduce the renormalization group equation
effectively in the space of the finite number of coupling constants
$\tilde g_k \;  (2\le k \le m)$
\begin{equation}
\left[N\frac{\partial}{\partial N}
- \sum_{k=2}^{m}
 \betaeff_k(\tg_2,\ldots,\tg_m)
\frac{\partial}{\partial \tg_k}
+ 1 \right]
 F(N,\tg_2,\ldots,\tg_m) = r( \tg_2,\ldots,\tg_m).
\end{equation}
The $m$-th multicritical point
should be obtained as a simultaneous zero
of all the beta functions
$\betaeff_2=\cdots=\betaeff_m=0$.
The susceptibility exponent is given by an eigenvalue of the
matrix of derivatives of beta functions
\be
\Omega_{ij}=\partial \betaeff_i/\partial \tg_j
\ee
at the fixed point,
which is an $(m-1) \times (m-1)$ real matrix.

To illustrate the multi-coupling case explicitly, we take
the case of two coupling constants.
We can solve the reparametrization identities recursively to
leading order in $1/N$, and combine
$\tilde \beta_k(\tilde g_2, \tilde g_3, \tilde g_k\!=\!0 \; (k\ge 4))$
to obtain the effective beta functions
\begin{eqnarray}
\betaeff_2(\tg_2,\tg_3)
&\!\!\! = &\!\!\! \frac{-2 \alpha\beta\gamma}
{(\alpha\beta + \beta\gamma +\gamma\alpha)^2}
\frac{(\beta^3 - \gamma^3)\log \alpha + \mbox{cyclic}}%
{(\alpha-\beta)(\beta-\gamma)(\gamma-\alpha)}+
\frac{\alpha + \beta + \gamma}{\alpha\beta + \beta\gamma
+\gamma\alpha}, \\
\betaeff_3(\tg_2,\tg_3)
&\!\!\! = &\!\!\! \frac{3 \alpha\beta\gamma}
{(\alpha\beta + \beta\gamma +\gamma\alpha)^2}
\frac{(\beta^2 - \gamma^2)\log \alpha + \mbox{cyclic}}%
{(\alpha-\beta)(\beta-\gamma)(\gamma-\alpha)}
- \frac{2}{\alpha\beta + \beta\gamma +\gamma\alpha},
  \label{multibeta}
\end{eqnarray}
where $\alpha,\beta,\gamma$ are the three roots of the cubic equation
\begin{equation}
  \tg_3 x^3 + \tg_2 x^2 + x -1 = 0.
 \label{eqn:saddlepointeq}
\end{equation}

We find three fixed points as the simultaneous zeros
of $\betaeff_2$ and $ \betaeff_3 $
(See figure 3).%~\ref{fig:2flow}).
% figure 3

The first fixed point is at
$(\tg_2, \tg_3)=(-1/3,1/27)$.
This result agrees with the exact value of the coupling
constants at the $m=3$ critical point.
We have evaluated the derivative matrix
$\Omega_{ij}=\partial \betaeff_i/\partial \tg_j$
by expanding the effective beta function around the fixed point
\begin{equation}
\Omega_{ij}
\left(\tilde g_2=-{1 \over 3},\tilde g_3={1 \over 27}\right)=
\left(
  \begin{array}{cc}
    \frac{11}{10} & \frac{9}{5} \\
    -\frac{7}{60} & \frac{3}{20}\\
  \end{array}
  \right).
\end{equation}
One of the eigenvalues of the derivative matrix is 3/4 which gives
the exact susceptibility exponent $\gamma_1 = 4/3$,
while the other eigenvalue is 1/2 which gives an analytic term.
The corresponding eigenvectors also agree with the exact solution.
As we stressed before, we can regard $N^{1/\gamma_1}$ as an
ultraviolet cut-off $\Lambda^2$ in taking the double scaling
limit.
Since both eigenvalues are positive,
this is an ultraviolet fixed point, or twice unstable fixed point.
This means that we have to tune two bare coupling constants
for taking the double scaling limit.

The second fixed point is at $\tg_2= -1/4$ and $\tg_3 = 0$.
We obtain the derivative matrix
$\Omega_{ij}=\partial \betaeff_i/\partial \tg_j$
at $(\tg_2,\tg_3)=(-\frac{1}{4},0)$
\begin{equation}
  \Omega_{ij}
 \left(\tilde g_2=-{1 \over 4},\tilde g_3=0\right)
  = \left(
  \begin{array}{cc}
    \frac{2}{3} & \frac{7}{3} \\ 0 & - \frac{1}{2}\\
  \end{array}
\right), \end{equation}
which has eigenvalues $2/3$ and $-1/2$.
The eigenvalue $2/3$ correctly gives the $m=2$ susceptibility exponent
$\gamma_1=3/2$.
The fixed point is infrared-repulsive in the direction
$(\delta \tg_2,  \delta \tg_3) \propto (1,0)$
and infrared-attractive in the direction
$(\delta \tg_2,  \delta \tg_3) \propto (-2,1)$,
which are the eigenvectors for the eigenvalues $2/3$ and $-1/2$,
respectively.
The infrared-attractive direction is tangential to the $m=2$ critical line
\begin{eqnarray}
\tg_3 & = & - \frac{1}{27} [ 2+ 9\tg_2 + 2( 1+ 3 \tg_2)^{3/2}],
  \label{eqn:critical_line1} \\
\tg_3 & = & - \frac{1}{27} [ 2+ 9\tg_2 - 2( 1+ 3 \tg_2)^{3/2}]\; \;
( \tg_2 <0).
  \label{eqn:critical_line2}
\end{eqnarray}
Moreover, a branch of the  $m=2$ critical line (\ref{eqn:critical_line1})
exhibits to be a trajectory of
the renormalization group flow approaching to $(0,-1/4)$ in the
infrared limit. This means that we have only to tune the
bare coupling constants on this line to achieve the $m=2$
critical point. The renormalized coupling constants can take
values between $(-\infty,0)$ and $(0,0)$ on the $\tg_2$ axis.

%The third fixed point is the trivial fixed point at the origin
%$(\tg_2, \tg_3) = (0,0)$.
The third fixed point is the trivial fixed point at the origin
$(\tg_2, \tg_3) = (0,0)$.

We admit that we do not understand the reason why tuning on the
other branch (\ref{eqn:critical_line2}) leads us to $m=2$ criticality.
Though the line (\ref{eqn:critical_line2}) is a trajectory of the
renormalization group flow connecting two fixed points,
the flow reaches the gaussian fixed point in the infrared limit.
This might be due to the fact that the scaling matrix at the origin
has a divergence of $\log \tg_3$ order.

We note that the $m=2$ fixed point and the trivial fixed  point are
in correspondence with the ones found in the analysis of the single
coupling constant case.
Actually we find that the $\tg_2$-axis is an invariant surface and
the effective beta function $\betaeff_2(\tg_2,0)$ on it is
identical to the effective beta function $\betaeff_2(\tg_2)$
in eq.(\ref{eqn:effectivebeta}) obtained in the single coupling
constant case.
This is not surprising. Two operations, setting $\tg_3$ to be
zero and solving the reparametrization identities recursively,
are commutative.

Now we consider the general case of the flow in the $m-1$
dimensional coupling constant space
$\Gamma_m = \{(\tg_2,\tg_3,\ldots,\tg_m,0,0,\ldots)\}.$
This space includes the subspace $\Gamma_k ( 2 \le k <m)$.
We find that
the effective beta functions for the additional coupling constants
vanish
 everywhere on $\Gamma_k$,
\be
\betaeff_{k+1}=0,\cdots, \betaeff_m=0, \quad
{\rm if} \quad  g_{k+1} =\cdots = g_m =0.
\ee
The effective renormalization group equation with $g_{k+1} =
\cdots = g_m =0$ turns out to be identical with that obtained in
the $k-1$dimensional case.
Therefore any fixed point in $\Gamma_k$ is a fixed point in
$\Gamma_m$ if $m>k$.

The behavior of the free energy near a
fixed point is determined by the derivative matrix $\Omega$ at that
point. We denote the $j$-th eigenvalue of $\Omega$ as $\lambda_j$ and
its eigenvector as $V_{jl}$
\be
\sum_{k=2}^m V_{jk}\Omega_{kl}=
\sum_{k=2}^m V_{jk}
{\partial \betaeff_{k} \over \partial \tilde g_l}
(\tilde g_*) =\lambda_j  V_{jl}.
\ee
The renormalization group equation tells us that the singular part
of the free energy is given by
\be
N F_{sing}(N,\tilde g)=\sum_{h=0}^{\infty}a_h
\biggl[N \sum_{j=2}^m
\bigl(\sum_{k=2}^m V_{jk}
(\tilde g_k-\tilde g_{k*})\bigr)^{1 \over \lambda_j}
\biggr]^{1-h}
\ee
To take a double scaling limit, fine tuning
of the coupling constants is necessary.
If we start near the fixed point and let $N$ decrease
(infrared direction),
only those coupling constants corresponding to
positive eigenvalues of $\Omega$ can survive.
Eventually, the combination of coupling constants corresponding to
the maximal eigenvalue dominates in this limit and the
susceptibility exponent $\gamma_1$ is given by the inverse of
this maximal eigenvalue.

Correlation functions among the operators
$N(\phi^2)^2,\ldots,N(\phi^2)^k$ can be dealt with in $\Gamma_m$
if $k \le m$.
These correlation functions are given by the
derivatives of the free energy $F$ with respect to the coupling
constants $\tg_2,\ldots,\tg_k$.
\be
\VEV{{-N \over 2k_1}(\phi^2)^{k_1} \cdots
{-N \over 2k_n}(\phi^2)^{k_n}}_c=
{\partial \over \partial \tilde g_{k_1}} \cdots
{\partial \over \partial \tilde g_{k_n}}
N F(N, \tilde g).
\ee
In taking the double scaling limit,
we can tune coupling constants by fixing all the scaling variables
\be
N\sum_k [V_{jk}(\tilde g_k-\tilde g_{k*})]^{1 \over \lambda_j},
\qquad j=2, \cdots, m
\ee
corresponding to the $j$-th eigenvalue $\lambda_j$ of the derivative
matrix $\Omega$.
Then the operators corresponding to these combinations of
coupling constants obey a scaling
law near a fixed point determined by the derivative matrix
$\Omega$. The scaling behavior of a single point function is
\be
\left\langle \sum_{k=2}^m V^{-1}_{kj}{N \over 2k} (\phi^2)^k \right\rangle
\sim N^{\lambda_j},
\ee
where $\lambda_j$ is its eigenvalue.
The operator belonging to a positive (negative) eigenvalue is a
relevant (irrelevant) operator and is amplified (deamplified)
as $N\rightarrow \infty$.

The exact result shows that the $m$-th multicritical point
exists in $\Gamma_m$ and is isolated from $\Gamma_k$ for $k <m$.
Fine tuning of the $m-1$ coupling constants is needed for the
double scaling limit at this critical point.
The knowledge of the exact solution in the vector model
\cite{NIYO}--\cite{ZJ}
helps us to conjecture the nature of the renormalization group
flow in the $m-1$ dimensional coupling constant space.
When we consider space $\Gamma_m$ of $m-1$ coupling constants,
we should obtain
the critical points by demanding
\be
\betaeff_2 = \cdots = \betaeff_m = 0.
\ee
The $k$-th multicritical point for $m \ge k \ge 2$ can be obtained
as a fixed point in this $m-1$ coupling constant space $\Gamma_m$.
For the $k$-th multicritical point,
the fine tuning of $k-1$ parameters is necessary to achieve the
double scaling limit.
Corresponding to these tuning parameters,
there are $k-1$ relevant operators which should be given by
linear combinations of $(\phi^2)^2,\ldots,(\phi^2)^k$.
All other operators $(\phi^2)^{k+1},\ldots,(\phi^2)^m$ should be
irrelevant near the critical point.
In other words, the $k$-th multicritical point in
$\Gamma_m (m > k)$ has $k-1$ repulsive directions and $m-k$
attractive directions.

%%%%%%% Discussion on the matrix models %%%%%%%%%%%%%%%%%%%%%%%%%%%%
\section {Discussions}
We have applied the renormalization group approach for large $N$
proposed by Br\'{e}zin and Zinn-Justin to vector models.  We
have found that we have to consider an effective renormalization
group flow in the finite-dimensional coupling constants space in
order to reproduce the exact critical coupling constants
and the exact critical exponents.
In the procedure, the reparametrization identities have played a
crucial role.
We have obtained explicit expressions of beta functions in the
case of a single coupling constant and two coupling constants.
We have found that the expressions obtained by approximate
evaluation, namely the $1/N$-expansion and the perturbative
expansion in the coupling constant, gives the result in complete
agreement with those obtained from the exact difference equations.

The phase structure for two-term potential has been clearly
understood in the language of the renormalization group.
As we described earlier, a naive application of the
renormalization group approach to matrix models has given us results
which show no improvement as we go to higher orders.
In view of our analysis of the exact
result on the flow of the coupling constants for
the vector model,
one should not be surprised by these results on matrix models.
We should look for an effective renormalization group flow by
taking account of the reparametrization freedom in the field space.
Although the above results are derived exactly only for the
vector model, we have a similar ambiguity to identify the
renormalization group flow in the coupling constant space for
matrix models.
For instance, there are identities which form a representation
of Virasoro algebra in matrix models \cite{FKN}.
There has been some work to interpret the Virasoro constraints as
reparametrization identities in the field space in the matrix
model \cite{ITOYAMAMATSUO}.
We are trying to devise a way to implement
our idea of the effective renormalization group flow through the
application of the reparametrization identities to matrix models
in order to make the renormalization group approach more useful.
Work along this line is in progress.
%
%%%%%%%%%%%%%%%%%%%%%%%%%%%%%%%%%%%%%%%%%%%%%%%%%%%%%%%%%%%%%
\par
\vspace{3mm}
We thank B.~Durhuus and T.~Hara for an illuminating discussion and
P.~Crehan for a careful reading of a part of the manuscript.
This work is supported in part by Grant-in-Aid for Scientific
Research (S.H.) and Grant-in-Aid for Scientific
Research for Priority Areas (No. 04245211) (N.S.) {}from
the Ministry of Education, Science and Culture.
\vspace{3mm}

\newpage
{\bf Figure captions}

\noindent
{\bf figure 1}
The effective beta function $\betaeff(\tg_2)$ and the
inhomogeneous term $r(\tg_2)$. The real parts are shown in
solid curves and the imaginary parts in dashed curves.

\noindent
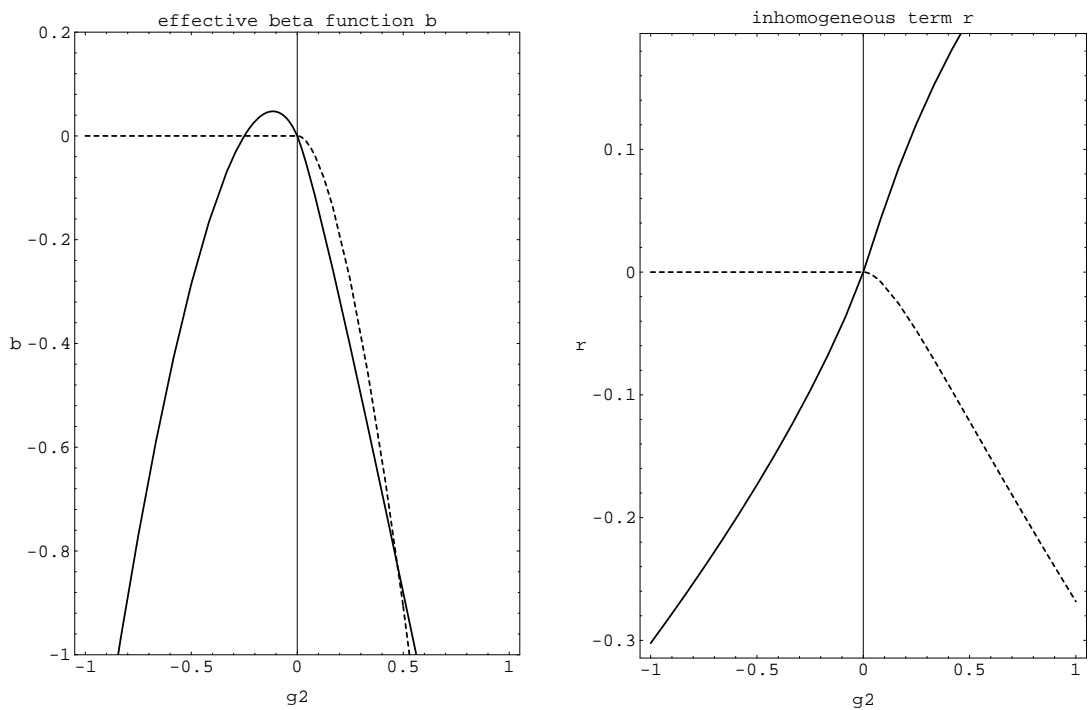
{\bf figure 2}
The
free energy in the leading order in $N$. The singular part
and analytic part around $\tg_2=-1/4$ are drawn separately.
The real part is drawn in solid curves and the imaginary
part in dashed curves.

\noindent
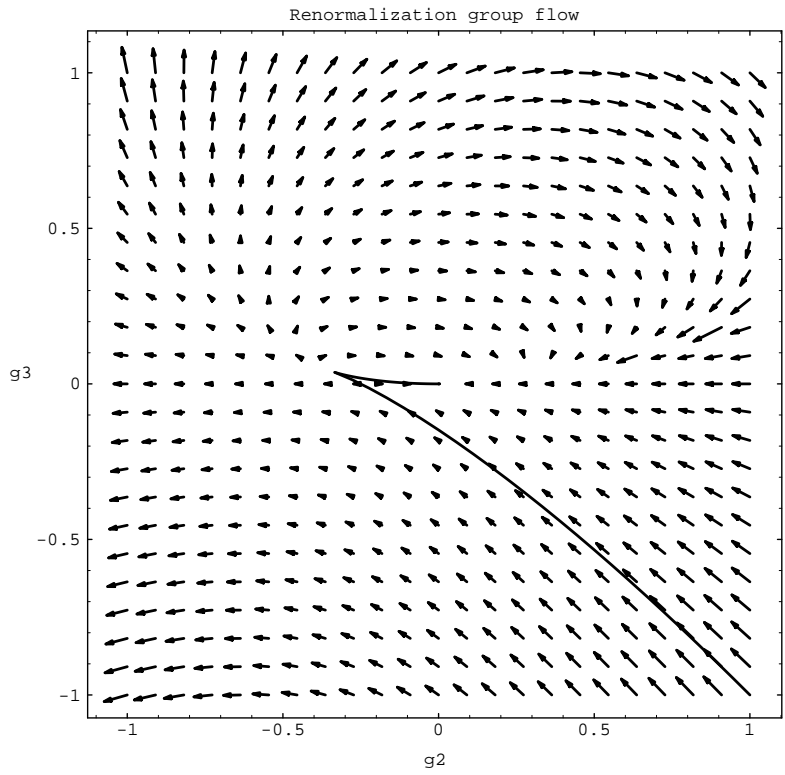
{\bf figure 3}
Renormalization group flow in the infrared
(decreasing $N$) direction. One shows the global
behavior of the flow, while the other shows the detail
of the flow near the origin. The $m=2$ critical line is
shown in a solied curve.

\begin{thebibliography}{10}
%
\bibitem{BIPZ} E.~Br\'ezin, C.~Itzykson, G.~Parisi and J.-B.~Zuber,
           {\it Commun.~Math.~Phys.~}{\bf 59} (1978) 59;
           D.~Bessis, C.~Itzykson and J.-B.~Zuber,
           {\it Adv.~Appl.~Math.~}{\bf 1} (1980) 109;
%.
%\bibitem{KAKOMI}
           V.A.~Kazakov, I.K.~Kostov and A.A.~Migdal,
           {\it Phys.~Lett.~}{\bf 157B} (1985) 295;
           F.~David, {\it Nucl.~Phys.~}{\bf B257} (1985) 45;
           V.A.~Kazakov and A.A.~Migdal,
           {\it Nucl.~Phys.~}{\bf B311} (1988) 171.
\bibitem{GRMI} D.J.~Gross and A.A.~Migdal,
           {\it Phys.~Rev.~Lett.~}{\bf 64} (1990) 127; 717,
           {\it Nucl.\ Phys.\ }{\bf B340} (1990), 333;
           E.~Br\'ezin and V.A.~Kazakov,
           {\it Phys.~Lett.~}{\bf 236B} (1990) 144;
           M.~Douglas and S.~Shenker,
           {\it Nucl.~Phys.~}{\bf B335} (1990) 635;
\bibitem{GMIL}
           D.J.~Gross and N.~Miljkovi\'c,
           {\it Phys.\ Lett.\ }{\bf 238B} (1990) 217;
           E.~Br\'ezin, V.A.~Kazakov and Al.B.~Zamolodchikov,
           {\it Nucl.\ Phys.\ }{\bf B338} (1990) 673;
           P.~Ginsparg and J.~Zinn-Justin,
           {\it Phys.\ Lett.\ }{\bf 240B} (1990) 333;
           G.~Parisi, {\it Phys.\ Lett.\ }{\bf 238B} (1990) 209, 213.
%
\bibitem{DDK} F.~David,
        {\it Mod.\ Phys.\ Lett.\ }{\bf A3} (1988) 1651;
        J.~Distler and H.~Kawai,
        {\it Nucl.\ Phys.\ }{\bf B321} (1989) 509;
        J.~Distler, Z.~Hlousek and H.~Kawai,
        {\it Int.\ J. Mod.\ Phys.\ }{\bf A5} (1990) 391; 1093.
\bibitem{BEKL} M.~Bershadsky and I.R.~Klebanov,
           {\it Phys.\ Rev.\ Lett.\ }{\bf 65} (1990) 3088;
            N.~Sakai and Y.~Tanii,
           {\it Int.\ J. Mod.\ Phys.\ }{\bf A6} (1991) 2743.
%
\bibitem{BRHI} E.~Br\'ezin and S.~Hikami,
           {\it Phys.\ Lett.\ }{\bf 283B} (1992) 203;
           {\it Phys.\ Lett.\ }{\bf 295B} (1992) 209.
%          Ecole Normale preprint LPTENS-92-31 (September 1992).
\bibitem{DDSW} S.R.~Das, A.~Dhar, A.M.~Sengupta and S.R.~Wadia,
        {\it Mod.\ Phys.\ Lett.\ }{\bf A5} (1990) 1041;
%\bibitem{AGBC}
        L.~Alvarez-Gaum\'{e}, J.L.F.~Barb\'{o}n and \v{C}.~Crnkovi\'{c},
        {\it Nucl.\ Phys.\ }{\bf B394} (1993) 383;
%        preprint CERN-TH-6600-92 (July 1992).
\bibitem{DFJ} B.~Durhuus, J.~Fr\"ohlich and T.~J\'onsson,
        {\it Nucl.\ Phys.\ }{\bf B240} (1984) 453,{\bf B257} (1985) 779;
        J.~Ambj{\o}rn, B.~Durhuus, and J.~Fr\"ohlich,
        {\it Nucl.\ Phys.\ }{\bf B257} (1985) 433.
%
\bibitem{BRZJ} E.~Br\'ezin and J.~Zinn-Justin,
        {\it Phys.\ Lett.\ }{\bf B288} (1992) 54.
\bibitem{WIKO} K.G.~Wilson and J.~Kogut,
        {\it Phys.\ Rep.\ }{\bf 12C} (1974) 75.
\bibitem{ALDA} J.~Alfaro and P.~Damgaard,
        {\it Phys.\ Lett.\ }{\bf B289} (1992) 342;
% .
%\bibitem{PERI}
        V.~Periwal,
        {\it Phys.\ Lett.\ }{\bf B294} (1992) 49;
%Princeton preprint IASSNS-HEP-92-47 (1992);
%\bibitem{GAO}
        H.~Gao,         Trieste preprint IC 302-92 (1992).
\bibitem{CARL} J.~Carlson,
        {\it Nucl.\ Phys.\ }{\bf B248} (1984) 536;
        R.~Brustein and S.P.~de~Alwis,
        University of Colorado preprint COLO-HEP-253 (1991).
%
\bibitem{NIYO} S.~Nishigaki and T.~Yoneya,
        {\it Nucl.\ Phys.\ }{\bf B348} (1991) 787;
        {\it Phys.\ Lett.\ }{\bf B268} (1991) 35.
\bibitem{DVKO} P.~Di~Vecchia, M.~Kato and N.~Ohta,
        {\it Nucl.\ Phys.\ }{\bf B357} (1991) 495;
        {\it Int.\ J. \ Mod.\ Phys.\ }{\bf 7A} (1992) 1391;
%\bibitem{DVM}
        P. Di Vecchia and M. Moshe,
        {\it Phys.\ Lett.\ }{\bf B300} (1992) 49.
%        Technion preprint TECHNION-PH-54-92 (1992).
\bibitem{AMP} A.~Anderson, R.~Myers and V.~Periwal,
        {\it Phys.\ Lett.\ }{\bf B254} (1991) 89;
        {\it Nucl.\ Phys.\ }{\bf B360} (1992) 463.
\bibitem{ZJ} J.~Zinn-Justin,
        {\it Phys.\ Lett.\ }{\bf B257} (1991) 335;
        Saclay preprint SPhT/91-054, 91-185 (1991).
%
\bibitem{HIS}
         S.~Higuchi, C.~Itoi and N.~Sakai, 
        {\it Phys.~Lett.~}{\bf 312B} (1993) 88;
%         Tokyo Institute of Technology preprint TIT/HEP-215, 
%         hep-th/9303090 (1993).
\bibitem{FKN} M.~Fukuma, H.~Kawai and  R.~Nakayama,
        {\it Int.~J.~Mod.~Phys.~}{\bf 6A} (1991) 1385;
        R.~Dijkgraaf, E.~Verlinde and H.~Verlinde,
       {\it Nucl.~Phys.~}{\bf B348} (1991) 435.
%        E.~Verlinde and H.~Verlinde,
%       {\it Nucl.~Phys.~}{\bf B348} (1991) 457.
\bibitem{ITOYAMAMATSUO}
         H.~Itoyama and Y.~Matsuo,
         {\it Phys.\ Lett.\ }{\bf B262} (1991) 233;
         L. Alvarez-Gaum\'{e}, C.~Gomez and J.~Lacki,
         {\it Phys.\ Lett.\ }{\bf B253} (1991) 56;
         A.~Mironov and A.~Morozov,
         {\it Phys.\ Lett.\ }{\bf B252} (1990) 47.
%
\end{thebibliography}
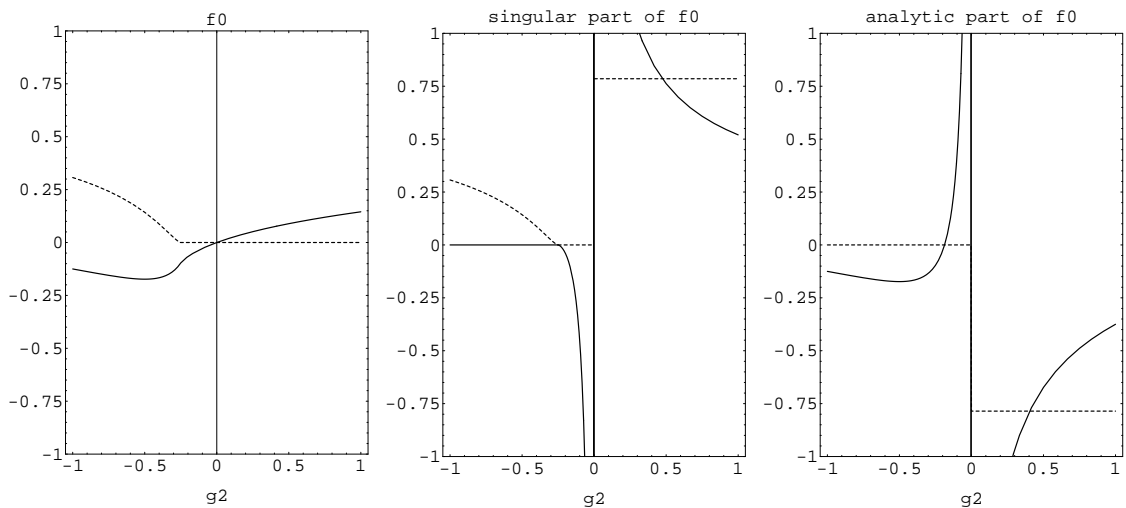
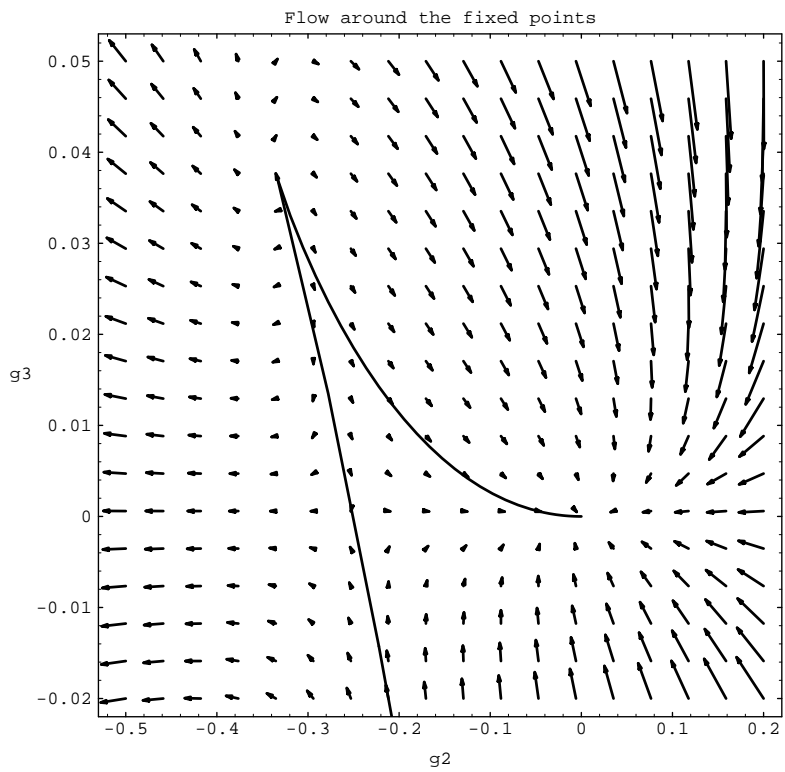
\end{document}